\documentclass[twocolumn,
showpacs,preprintnumbers,amsmath,amssymb]{revtex4}

\usepackage{amssymb,amsmath,amsthm}
\usepackage{epsfig}
\usepackage{multirow}

\theoremstyle{definition}

\newcommand{\bra}[1]{{\left\langle #1 \right|}}
\newcommand{\ket}[1]{{\left| #1 \right\rangle}}

\newcommand{\T}{\mbox{$\mathrm{tr}$}}

\begin{document}
\title{General polygamy inequality of multi-party quantum entanglement}

\author{Jeong San Kim}
\email{freddie1@suwon.ac.kr} \affiliation{
 Department of Mathematics, University of Suwon, Kyungki-do 445-743, Korea
}
\date{\today}

\begin{abstract}
Using entanglement of assistance, we establish a general polygamy inequality
of multi-party entanglement in arbitrary dimensional quantum systems.
For multi-party closed quantum systems, we relate our result with the
monogamy of entanglement to show
that the entropy of entanglement is an universal entanglement measure
that bounds both monogamy and polygamy of multi-party quantum entanglement.
\end{abstract}

\maketitle

Entanglement has been recognized as a key resource in many quantum
information processing tasks such as quantum
teleportation~\cite{tele} and dense coding~\cite{BW92}. Entanglement
can be used to perform quantum key distribution where
entangled-state analysis is used to prove the security of privacy
amplification~\cite{BB84,Eke91}. Entanglement is also essential in
certain models of quantum computing such as non-deterministic gate
teleportation~\cite{NC+97} and one-way quantum
computation~\cite{RB01}.

One distinct phenomenon of quantum entanglement from classical
correlation is that it cannot be shared freely in multi-party
systems. A simple example arises when a pair of parties in a
multi-party quantum system share maximal entanglement, in which case
they cannot have any entanglement nor classical
correlations with other parties. This restricted shareability of
entanglement in multi-party quantum systems is known as the {\em
Monogamy of Entanglement} (MoE)~\cite{T04}. MoE does not have any
classical counterpart because all classical probability
distributions can be shared among parties, and correlation between a
pair of parties (whether they are perfectly correlated or not) does
not restrict other parties' correlation. Thus MoE makes quantum
physics fundamentally different form classical physics.

The first mathematical characterization of MoE was established by
Coffman-Kundu-Wootters (CKW) for three-qubit systems using tangle as
the bipartite entanglement measure~\cite{CKW}. For a three-qubit
state $\rho_{ABC}$ with two-qubit reduced density matrices
$\T_C\ket{\psi}_{ABC}\bra{\psi}=\rho_{AB}$ and
$\T_B\ket{\psi}_{ABC}\bra{\psi}=\rho_{AC}$,
\begin{equation}
\tau\left(\rho_{A(BC)}\right)\geq \tau\left(\rho_{AB}\right)+\tau\left(\rho_{AC}\right),
\label{eq: CKW}
\end{equation}
where $\tau\left(\rho_{A(BC)}\right)$ is the entanglement of
$\rho_{ABC}$ with respect to the bipartition between $A$ and $BC$
measured by tangle, $\tau\left(\rho_{AB}\right)$ and
$\tau\left(\rho_{AC}\right)$ are tangle of $\rho_{AB}$ and
$\rho_{AC}$ respectively. Inequality~(\ref{eq: CKW}) (also referred
as CKW inequality) shows the mutually exclusive nature of
multi-party quantum entanglement in a quantitative way; more
entanglement shared between two qubits ($A$ and $B$) necessarily
implies less entanglement between the other two qubits ($A$ and
$C$).
Later, CKW inequality was generalized for multi-qubit systems rather 
than just three qubits~\cite{OV}.

However, CKW inequality is know to fail in its generalization for
higher-dimensional quantum systems due to the existence of quantum
states violating Inequality~(\ref{eq: CKW})~\cite{KDS}. Moreover,
this characterization of MoE in forms of an inequality is not
generally true even in three-qubit systems for other entanglement measures;
one can easily find an example of
three-qubit state that violates CKW inequality if we use Entanglement
of Formation~(EoF)~\cite{BDSW} instead
of tangle. Thus having a proper way of quantifying bipartite
entanglement is crucial in the study of MoE.

Later, monogamy of multi-qubit
entanglement and some cases of higher-dimensional quantum systems
were characterized in terms of various entanglement
measures~\cite{KDS, KSRenyi, KT, KSU}. For general monogamy
inequality of multi-part entanglement, it was recently shown that
squashed entanglement~\cite{CW04} is a faithful entanglement measure
(it vanishes only for separable states)~\cite{BCY10}, which also shows
monogamy inequality of multi-party entanglement in arbitrary
dimensional quantum systems~\cite{KW}.

Whereas MoE is about the restricted shareability of bipartite
entanglement in multi-party quantum systems, the dual concept of
bipartite entanglement namely {\em Entanglement of Assistance} (EoA) is
known to have a dually monogamous (thus polygamous) property in
multipartite quantum systems; for a three-qubit pure state
$\ket{\psi}_{ABC}$, {\em Polygamy of Entanglement} (PoE) was
characterized as a dual inequality~\cite{GMS, GBS}
\begin{equation}
\tau\left(\ket{\psi}_{A(BC)}\right)\le\tau_a\left(\rho_{AB}\right)
+\tau_a\left(\rho_{AC}\right), \label{3dual}
\end{equation}
where $\tau_a\left(\rho_{AB}\right)$ and
$\tau_a\left(\rho_{AC}\right)$ are the tangle of
assistance~\cite{GBS} of $\rho_{AB}$ and $\rho_{AC}$ respectively.

For MoE characterized by CKW inequality, the bipartite entanglement
between $A$ and $BC$ measured by tangle is an upper bound for the
sum of two-qubit entanglement between $A$ and each of $B$ and $C$.
Interestingly, the same quantity also plays as a lower bound for the
sum of two-qubit entanglement of assistance in the polygamy
inequality. Later PoE was generalized into multi-qubit
systems~\cite{GBS, KT} and three-party pure states of arbitrary
dimension~\cite{BGK}. Recently, a tight upper bound on polygamy
inequality was also proposed in an arbitrary-dimensional
multi-party quantum systems~\cite{Kim09}.
However, a general polygamy inequality of
multi-party, higher-dimensional quantum systems was an open
question.

The study of quantum entanglement in higher-dimensional quantum
systems is important for not only theoretical sense but
practical reasons as well; MoE can restrict the possible correlation
between authorized users and the eavesdropper, which tightens
security bounds in quantum cryptography (QC).
Furthermore, to optimize the efficiency of entanglement usage
as a resource in QC, higher-dimensional quantum systems rather than
qubits are preferred in some physical systems for stronger security in
quantum key distribution (QKD)~\cite{GJV+06}.
However, generalization from qubits to qudits is not straightforward
for example, in the complexity of a no-go theorem for
universal transversal gates in
quantum error correlation~\cite{CCCZC08}.

Here, we provide a polygamy inequality of entanglement that holds for
multi-party quantum systems of arbitrary high dimensions.
For multi-party closed quantum systems, we relate our result with the
monogamy inequality~\cite{KW}, and show
that the entropy of entanglement serves as both upper and lower bounds
for monogamy and polygamy of multi-party quantum entanglement. Thus
the entropy of entanglement is an universal entanglement measure
that bounds both MoE and PoE.

Let us recall the definition of EoF and EoA of bipartite quantum
systems. For a bipartite pure state $\ket{\psi}_{AB}$, its
entropy of entanglement is
\begin{align}
E\left(\ket{\psi}_{AB}\right):=S\left(\rho_A\right) \label{pureE}
\end{align}
where $\rho_A=\T_B \ket{\psi}_{AB}\bra{\psi}$ and $S\left(\rho
\right)=-\T \rho \log \rho$ is von Neumann entropy. For a mixed
state $\rho_{AB}$, its EoF is defined as the minimum average
entanglement
\begin{equation}
E_f\left(\rho_{AB}\right)=\min \sum_{i}p_i
E\left(\ket{\psi}_{AB}\right), \label{EoF}
\end{equation}
where the minimization is taken over all possible pure-state
decompositions of $\rho_{AB}=\sum_{i} p_i
\ket{\psi^i}_{AB}\bra{\psi^i}$. This procedure of minimizing over
all pure-state decompositions to determine mixed-state entanglement
is known as the convex-roof extension. From the convex-roof nature inherent
in the definition, EoF of~$\rho_{AB}$ is considered as the minimum amount
of entanglement needed to prepare $\rho_{AB}$, hence the terminology
{\em formation}.

As a dual quantity to EoF,
EoA of $\rho_{AB}$ is defined as the maximum average entanglement
\begin{equation}
E_a\left(\rho_{AB}\right)=\max \sum_{i}p_i
E\left(\ket{\psi}_{AB}\right), \label{EoA}
\end{equation}
over all possible pure-state decompositions of $\rho_{AB}$~\cite{LVvE03}.
If we consider $\rho_{AB}$ together with a purification
$\ket{\psi}_{ABC}$ such that $\rho_{AB}=\T
\ket{\psi}_\text{ABC}\bra{\psi}$, the party $C$ having the
purification of $\rho_{AB}$ can help increasing the entanglement of
$\rho_{AB}$ by performing measurements on its own system $C$ and
communicating the measurement results to $A$ and $B$.

Furthermore, the one-to-one correspondence between rank-one
measurements of $C$ and pure state ensembles of~$\rho_{AB}$ makes
this maximum possible average entanglement between $A$ and $B$ with
the assistance of $C$ as an intrinsic definition; the maximum
average entanglement over all possible pure-state decompositions of
$\rho_{AB}$, which is the definition of EoA in Eq.~(\ref{EoA}).
Thus, not only the mathematical duality between EoF and EoA  (one
takes the minimum whereas the other one takes the maximum), they
also have physical interpretations that are dual to each other; one
is the concept of formation and the other is the possible achievable
entanglement with assistance of the environment.

Before we discuss polygamy relation of multi-party entanglement in
terms of EoA, let us first consider a trade-off relation between EoA
and {\em one-way unlocalizable entanglement} (UE) in three-party
quantum systems~\cite{BGK}; for a three-party pure state
$\ket{\psi}_{ABC}$ with reduced density matrices
$\rho_{AB}$ and $\rho_{AC}$, we have
\begin{align}
S(\rho_A)&=E_u^{\leftarrow}(\rho_{AB})+E_a(\rho_{AC}),
\label{eq: 3UEEA}
\end{align}
where $E_u^{\leftarrow}(\rho_{AB})$ is UE of $\rho_{AB}$
\begin{equation}
E_u^{\leftarrow}(\rho_{AB}):=\min_{\{M_x\}} \left[S(\rho_A)-\sum_x p_x S(\rho^x_A)\right],\\
\label{eq:UE}
\end{equation}
with the minimum being taken over all possible rank-1 measurements
$\{M_x\}$ applied on system $B$.

Eq.~(\ref{eq: 3UEEA}) implies that
the entropy of entanglement of $\ket{\psi}_{ABC}$ with respect to the
bipartition between $A$ and $BC$ consists of two distinct parts: one is the
robust entanglement that can be localized onto $AC$ after the local
measurement of $B$ (denoted by $E_a(\rho_{AC})$), and the other part
is too sensitive to be localized onto $AC$ (denoted by
$E_u^{\leftarrow}\left(\rho_{AB}\right)$).

\begin{figure}
\includegraphics[width=7cm]{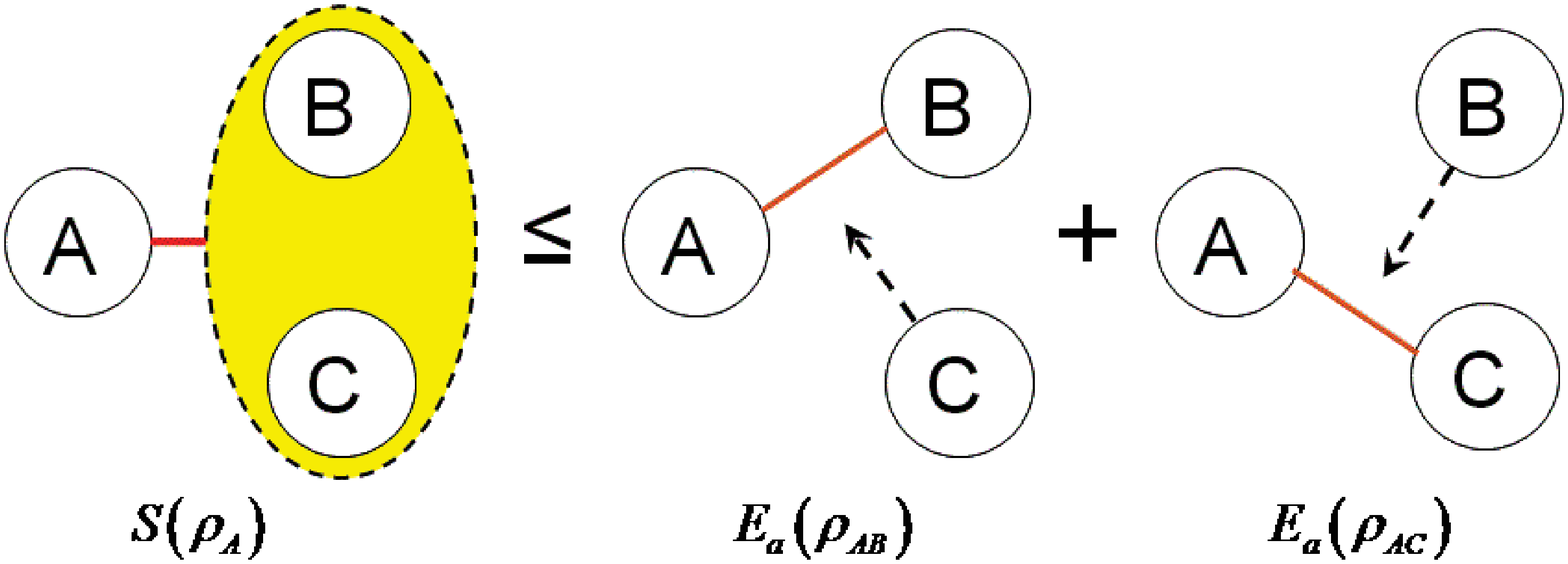}\\
\caption{The entanglement between $A$ and $BC$ of a three-party pure state $\ket{\psi}_{A(BC)}$ measured by $S(\rho_A)$
is always bounded by the sum of localizable entanglement between $A$ and $B$ measured by $E_a(\rho_{AB})$,
and between $A$ and $C$ measured by $E_a(\rho_{AC})$.}\label{fig1}
\end{figure}

UE is known to be subadditive under tensor product of quantum states,
and bounded below by the coherent information~\cite{BGK}.
Furthermore, form the quantitative relation between UE and mutual information
\begin{equation}
E_u^{\leftarrow}(\rho_{AB})\leq\frac{I(\rho_{AB})}{2}, \label{upper}
\end{equation}
Eq.~(\ref{eq: 3UEEA}) was shown to imply a trade-off relation
of localizable entanglement measured by EoA in three-party quantum systems;
for any tripartite pure state $\ket{\psi}_{ABC}$,
\begin{equation}
S(\rho_A)\leq E_a(\rho_{AB})+E_a(\rho_{AC}). \label{3poly}
\end{equation}
Thus, for a tripartite pure state of arbitrary dimension, there
always exists a polygamy relation of localizable entanglement that
can be quantified by the entropy of entanglement and EoA, which is
illustrated in Figure~\ref{fig1}.

Now we generalize Inequality~(\ref{3poly}) for arbitrary mixed
states of multi-party quantum systems rather than just three
parties. We first show that Inequality~(\ref{3poly}) is true for
tripartite mixed states, and use the result to establish a general
polygamy inequality of EoA in arbitrary dimensional multi-party
quantum systems. For a three-party mixed state $\rho_{ABC}$, let
$\rho_{A(BC)}=\sum_j p_j \ket{\psi_j}_{A(BC)}\bra{\psi^j}$ be an
optimal decomposition for EoA with respect to the bipartition
between $A$ and $BC$,
\begin{align}
E_a\left(\rho_{A(BC)}\right)=\sum_j p_j
E\left(\ket{\psi^j}_{A(BC)}\right), \label{3mixopt}
\end{align}
where $E\left(\ket{\psi^j}_{A(BC)}\right)=S\left(\rho^j_A \right)$
is the pure-state entanglement of $\ket{\psi^j}_{A(BC)}$ for each $j$ with
$\rho^j_A=\T_{BC}\ket{\psi^j}_{ABC}\bra{\psi^j}$.

For each $j$, $\ket{\psi^j}_{ABC}$ is a tripartite  pure state, therefore Inequality~(\ref{3poly}) leads us to
\begin{align}
E\left(\ket{\psi^j}_{A(BC)}\right) =& S\left(\rho^j_A \right)\nonumber\\
\leq& E_a\left(\rho^j_{AB}\right)+E_a\left(\rho^j_{AC}\right) \label{unipolypsii}
\end{align}
with $\rho^j_{AB}=\T_C \ket{\psi^j}_{ABC}\bra{\psi^j}$ and $\rho^j_{AC}=\T_B \ket{\psi^j}_{ABC}\bra{\psi^j}$.
The linearity of partial trace implies
\begin{align}
\sum_{j}p_j\rho_{AB}^j=\rho_{AB},~\sum_{j}p_j\rho_{AC}^j=\rho_{AC},
\label{rhosum}
\end{align}
and together with the definition of EoA, we have
\begin{align}
\sum_{j}p_jE_a\left(\rho_{AB}^j\right)&\leq E_a\left(\rho_{AB}\right),\nonumber\\
\sum_{j}p_jE_a\left(\rho_{AC}^j\right)&\leq E_a\left(\rho_{AC}\right).
\label{rhosum}
\end{align}
From the inequalities~(\ref{unipolypsii}), (\ref{rhosum}) and Eq.~(\ref{3mixopt}),
we thus have
\begin{align}
E_a\left(\rho_{A(BC)}\right)=&\sum_j p_j
E\left(\ket{\psi^j}_{A(BC)}\right)\nonumber\\
\leq &\sum_j p_jE_a\left(\rho^j_{AB}\right)+\sum_j
p_jE_a\left(\rho^j_{AC}\right) \nonumber\\
\leq& E_a\left(\rho_{AB}\right)+E_a\left(\rho_{AC}\right). \label{3polymixed}
\end{align}
In other words, Inequality~(\ref{3poly}) can be generalized for
tripartite mixed states in terms of EoA.

Now let us consider a multi-party quantum state $\rho_{A_1A_2\cdots
A_n}$ rather than just three parties. By letting $A_1=A$, $A_2=B$
and $A_3\cdots A_n=C$, we can consider $\rho_{A_1A_2\cdots A_n}$ as
a tripartite quantum state, and Inequality~(\ref{3polymixed}) leads
us to
\begin{align}
E_a\left(\rho_{A_1(A_2\cdots A_n)}\right)
\leq& E_a\left(\rho_{A_1A_2}\right)+E_a\left(\rho_{A_1(A_3\cdots A_n)}\right), \label{polymixed1}
\end{align}
where $\rho_{A_1A_2}=\T_{A_3\cdots A_n}\rho_{A_1A_2\cdots A_n}$,
$\rho_{A_1A_3\cdots A_n}=\T_{A_2}\rho_{A_1A_2\cdots A_n}$, and
$E_a\left(\rho_{A_1(A_3\cdots A_n)}\right)$ is EoA of
$\rho_{A_1A_3\cdots A_n}$ with respect to the bipartition between
$A_1$ and $A_3\cdots A_n$.
Because $\rho_{A_1A_3\cdots A_n}$ in Inequality~(\ref{polymixed1}) is a
$(n-1)$-party quantum state, we can apply Inequality~(\ref{3polymixed})
to obtain
$E_a\left(\rho_{A_1(A_3\cdots A_n)}\right) \leq
E_a\left(\rho_{A_1A_3}\right)+E_a\left(\rho_{A_1(A_4\cdots
A_n)}\right)$. Thus, by iterating
Inequality~(\ref{3polymixed}) on the last term of
Inequality~(\ref{polymixed1}), we obtain the following polygamy
inequality of multi-party entanglement
\begin{align}
E_a\left(\rho_{A_1(A_2\cdots A_n)}\right)
\leq& E_a\left(\rho_{A_1A_2}\right)+\cdots +E_a\left(\rho_{A_1A_n}\right). \label{npolymixed}
\end{align}
In contrast to monogamy inequality, which provides an upper bound on
the shareability of bipartite entanglement in multi-party quantum
systems, polygamy inequality in (\ref{npolymixed}) provides a
lower bound of how much entanglement can be created on bipartite
subsystems with assistance of the other parties.

Here we further note that
this upper and lower bounds of entanglement distribution in
multi-party quantum systems can be determined by a single quantity,
the entropy of entanglement for closed systems. The result about
three-party monogamy inequality in terms of
squashed entanglement (SE) in~\cite{KW} can easily be generalized
into an arbitrary multi-party quantum system; for a $n$-party state
$\rho_{A_1A_2\cdots A_n}$,
\begin{align}
E_{sq}\left(\rho_{A_1(A_2\cdots A_n)}\right)\geq E_{sq}\left(\rho_{A_1A_2}\right)+
\cdots +E_{sq}\left(\rho_{A_1A_2}\right),
\label{nmonomix}
\end{align}
where $E_{sq}\left(\rho_{AB}\right)$ is SE of $\rho_{AB}$ defined as
\begin{equation}
\label{eq:squashed}
E_{sq}\left(\rho_\text{AB}\right):=\frac{\inf\left\{S(\rho_{AE})
+S(\rho_{BE})-S(\rho_{ABE})-S(\rho_{E})\right\}}{2}
\end{equation}
with the infimum taken over all possible extension $\rho_{ABE}$ of
$\rho_{AB}$ such that $\T_E \rho_{ABE}=\rho_{AB}$. The quantity
inside the parenthesis of Eq.~(\ref{eq:squashed}) is quantum
conditional mutual information of $\rho_{ABE}$ denoted by
$I(A;B|E)$.

For a bipartite pure state~$\ket{\psi}_{AB}$, any possible extension
$\rho_{ABC}$ such that $\T_{C} \rho_{ABC}=\ket{\psi}_{AB}\bra{\psi}$
must be a product state $\ket{\psi}_{AB}\bra{\psi}\otimes\rho_{C}$
for some $\rho_{C}$ of subsystem $C$. From this fact, it is also
straightforward to verify that SE in~Eq.~(\ref{eq:squashed})
coincides with $S(\rho_{A})$ for any pure state $\ket{\psi}_{AB}$
with reduced density matrix $\rho_\text{A}$. In other words, for a
multi-party closed quantum system described by a pure state $\ket{\psi}_{A_1A_2\cdots A_n}$,
the monogamy inequality in terms of SE in~(\ref{nmonomix}) becomes
\begin{align}
S\left(\rho_{A_1}\right)\geq E_{sq}\left(\rho_{A_1A_2}\right)+ \cdots
+E_{sq}\left(\rho_{A_1A_2}\right), \label{nmonopure}
\end{align}
where $S\left(\rho_{A_1}\right)=E\left(\ket{\psi}_{A_1(A_2\cdots
A_n)}\right)$ is the entropy of entanglement of the pure state
$\ket{\psi}_{A_1A_2\cdots A_n}$ with respect to the bipartition
between
 $A_1$ and the other parties. Furthermore, from the definition of EoA in Eq.~(\ref{EoA}),
we also note that the left-hand side of polygamy inequality in~(\ref{npolymixed})
becomes the entropy of entanglement,
\begin{align}
S\left(\rho_{A_1}\right) \leq& E_a\left(\rho_{A_1A_2}\right)+\cdots
+E_a\left(\rho_{A_1A_n}\right), \label{npolypure}
\end{align}
for this closed quantum system described by
$\ket{\psi}_{A_1A_2\cdots A_n}$. Thus the entropy of entanglement
quantifying bipartite pure-state entanglement is an universal
entanglement measure that provides bounds for both monogamy and
polygamy of multi-party quantum entanglement.

To summarize, we have shown the polygamous nature of bipartite
entanglement distribution in multipartite quantum systems; using
EoA, we establish a general polygamy inequality of multi-party
entanglement in arbitrary high-dimensional quantum systems rather
than just qubits. For multi-party closed quantum systems, we have
related our polygamy inequality with the monogamy inequality in
terms of SE, and clarified that the entropy of entanglement serves
as both upper and lower bounds for MoE and PoE.

Our result completely characterizes the polygamous nature of
entanglement distribution in multi-party quantum systems of
arbitrary high dimension.
Noting the importance of the study on high-dimensional multipartite
entanglement, our result can provide a rich reference for future
work on the study of multipartite entanglement.

\section*{Acknowledgments}
This work was supported by Emerging Technology R\&D Center of SK Telecom.


\end{document}